\documentclass[a4paper]{article}

\usepackage{INTERSPEECH_v2}
\usepackage{graphicx}
\title{AV Speech Enhancement Challenge using a Real Noisy Corpus}
\name{Mandar Gogate$^1$, Ahsan Adeel$^2$, Kia Dashtipour$^3$, Peter Derleth$^4$, Amir Hussain$^1$$\dagger$}
\address{
$^1$Edinburgh Napier University, UK \\ 
  $^2$University of Wolverhampton, UK\\
  $^3$University of Stirling, UK \\ 
  $^4$Sonova AG, Switzerland 
}

\email{a.hussain@napier.ac.uk}
\begin{document}

\maketitle
\begin{abstract}
This paper presents, a first of its kind, audio-visual (AV) speech enhacement challenge in real-noisy settings. A detailed description  of the AV challenge, a novel real noisy AV corpus (ASPIRE), benchmark speech enhancement task, and baseline performance results are outlined. The latter are based on training a deep neural architecture on a synthetic mixture of Grid corpus and ChiME3 noises (consisting of bus, pedestrian, cafe, and street noises) and testing on the ASPIRE corpus. Subjective evaluations of five different speech enhancement algorithms (including SEAGN, spectrum subtraction (SS) , log-minimum mean-square error (LMMSE), audio-only CochleaNet, and AV CochleaNet) are presented as baseline results. The aim of the multi-modal challenge is to provide a timely opportunity for comprehensive evaluation of novel AV speech enhancement algorithms, using our new benchmark, real-noisy AV corpus and specified performance metrics. This will promote AV speech processing research globally, stimulate new ground-breaking multi-modal approaches, and attract interest from companies, academics and researchers working in AV speech technologies and applications. We encourage participants (through a challenge website sign-up) from both the speech and hearing research communities, to benefit from their complementary approaches to AV speech in noise processing.

\end{abstract}

\section{Introduction}
\label{s:introduction}
Evaluation of future AV speech enhancement technology requires real-world realistic AV corpora. Although there exist a number of small well-controlled AV speech corpora, such as BANCA \cite{bailly2003banca}, AVICAR \cite{lee2004avicar}, VidTIMIT \cite{sanderson2002vidtimit}, and Grid \cite{cooke2006audio}, but there is a need for evaluation of multi-modal speech enhancement systems using realistic audiovisual speech data. Specifically, AV datasets are required in which speakers are speaking more naturally than in many existing corpora, including conversational speech and imperfect visual data. This is represented by the speaker moving their head, obscuring their face, and also different levels of background noise to take account of the Lombard effect (where speakers naturally adjust their speech to take account of different levels of background noise). To our knowledge, there is no corpus available that contains a sufficient range of AV speech data, or variety of A and V noises (i.e., acoustic noise, speaker movement and occlusion, etc.).

ASPIRE, is a first of its kind high quality AV binaural speech corpus, recorded in real noisy settings such as cafeteria and restaurant. It is to be noted that, most of the existing AV SE methods use synthetic mixture of clean speech and noises for model evaluation. However, the synthetic mixture do not reflect the real noisy mixtures as speech is often reverberantly mixed with multiple competing noise background sources. Therefore, the ASPIRE corpus can be used by speech and machine learning communities as a benchmark resource to support reliable evaluation of AV SE technologies.

\begin{table}[!t]
  \caption{Grid Corpus Sentence Structure}
  \label{tab:Grid}
  \centering
  \scalebox{0.9}{
  \begin{tabular}{cccccc}
\hline    \multicolumn{1}{c}{\textbf{command}} &
  \multicolumn{1}{c}{\textbf{colour}} &
  \multicolumn{1}{c}{\textbf{preposition}} &
  \multicolumn{1}{c}{\textbf{letter}} &
  \multicolumn{1}{c}{\textbf{digit}} &
  \multicolumn{1}{c}{\textbf{adverb}}\\
    \hline
bin & blue & at & A-Z & 1--9 & again    \\
lay & green & by & minus W & zero & now    \\
place & red & in &  & & please    \\
set & white & with &  & & soon    \\

    \hline
  \end{tabular}
}
\end{table}

\section{ASPIRE Corpus}
\subsection{Sentence design}
ASPIRE corpus follows the same sentence format as the AV Grid corpus \cite{cooke2006audio} as shown in Table 1. The six words sentence consists of command, colour, preposition, letter, digit and adverb. The letter "w" was excluded because it is the only multi-syllabic letter. Each speaker produced all combinations of colour, letter and digit leading to 1000 utterances per talker in both real noisy settings and acoustically isolated booth. Thus, each talker recorded 2000 utterances.

\subsection{Speaker population}
Five speakers (two male and three female) contributed to the corpus. The speakers age ranged from 23 to 55. All the speakers have spent most of their lives in the United Kingdom and together encompassed a range of mixed English accents. All the participants were paid for their contribution. The corpus consists of total 10000 utterances (5000 recorded in real noisy settings, 5000 in acoustically isolated booth).

\begin{figure*}[!t]
  \centering
  \includegraphics[width=0.9\linewidth]{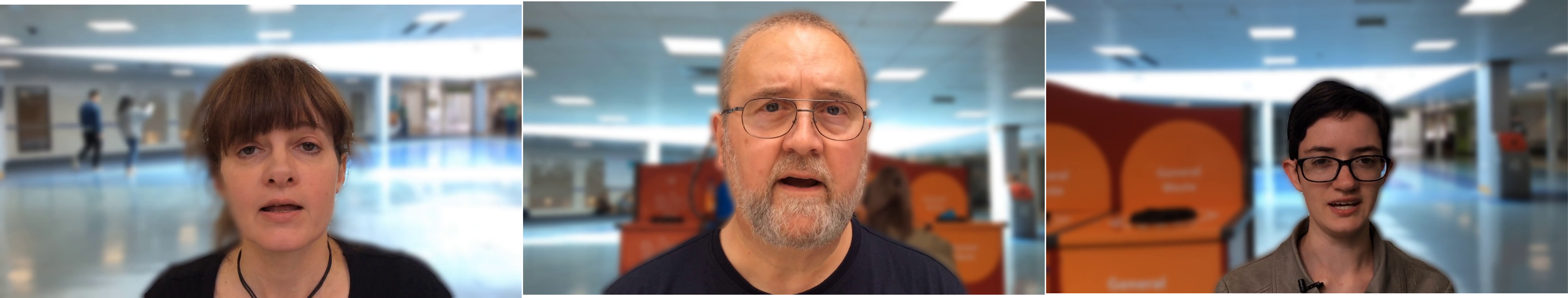}
  \caption{Sample video frames from ASPIRE corpus}
  \label{fig:screenshotsASPIRE}
\end{figure*}

\subsection{Collection} 
The dataset is recorded at the University of Stirling cafeteria and restaurant during busy lunch times (11.30 to 1.30) as well as in an acoustically isolated booth. The recording setup is shown in Figure 2. 

\begin{itemize}
\item Apple iPad-mini2 was used to record the video at 30 frames per second (fps) and 1080p resolution. It was placed at an eye level to avoid noise and distraction from the video apparatus. The distance between iPad and speaker was 90 centimetres. A collar microphone was also connected to the iPad. 
\item The high quality binaural audio from speaker is recorded using Zoom H4n pro recorder at a sampling rate of 44100 Hz and binaural microphone. The listener was wearing the binaural microphone at an approximate distance of 140 centimetres. The listener and speaker were sitting opposite to each other on the fixed chairs. 
\item Speakers were initially trained with a few utterances and were aware of the purpose of research. Periodic breaks were given to the speakers during recording to avoid fatigue. Speakers were instructed to read each sentence correctly without any interruption. Sentences were presented to the speaker on a laptop in a random order. Speakers were allowed to repeat the sentence in case the recording was interrupted or sentence was incorrectly uttered. In addition, the speakers were asked to repeat the utterance if the listener spotted any mistake. In 2000 utterances per speaker, around 2\% and 4\% of the utterances were re-recorded in booth and real noisy settings, respectively.

\end{itemize}

\begin{figure}[!t]
  \centering
  \includegraphics[width=\linewidth]{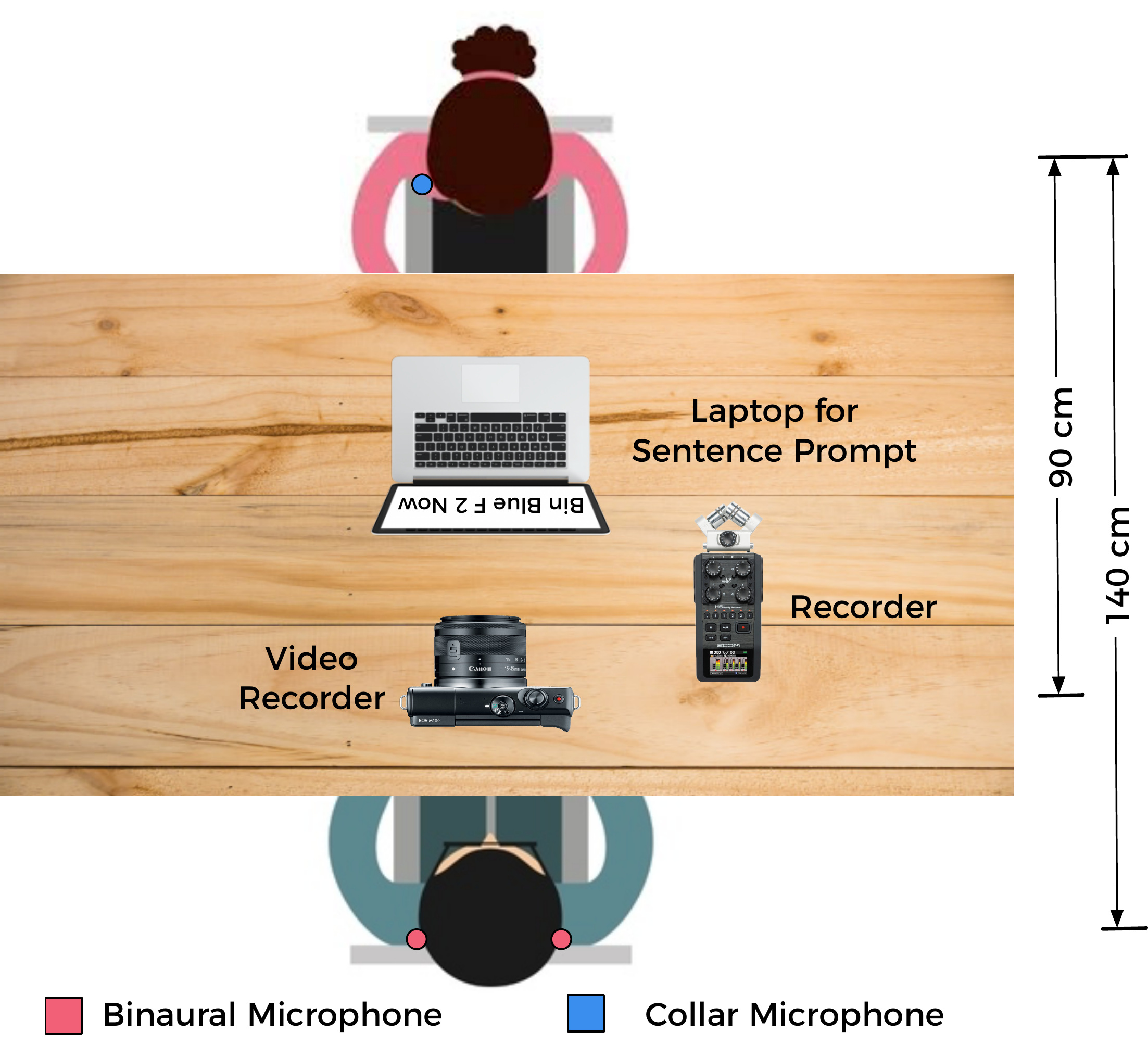}
  \caption{ASPIRE recording setting showing the location of listener, speaker, audio recorder, video recorder, sentence prompter and binaural/collar microphone}
  \label{fig:setupASPIRE}
\end{figure}

\section{Postprocessing}
\subsection{Audio postprocessing} 
The drift between audio and video data was calculated by synchronising the claps. The utterance start and end times were identified using Gentle (i.e. a forced-aligner built in Kaldi \cite{povey2011kaldi}), speech recorded from the collar microphone, and the transcript. Finally, all the segmented utterances were manually checked to correct any additional alignment errors. 
\subsection{Video postprocessing} 
The recorded raw videos had some clearly identifiable people other than the target speaker itself. Therefore, to ensure privacy, we estimated the speaker area for the first frame using a segmentation model and then pixelated the non-speaker area for the complete utterance using the estimated segmentation mask. This was possible because the speaker had same position throughout the utterance. Figure \ref{fig:screenshotsASPIRE} shows some sample video frames from the ASPIRE corpus.

\section{Baseline results}
\subsection{Baseline Speech Enhancement Framework}
The challenge uses CochleaNet \cite{m2019cochleanet} as a baseline AV speech enhancement approach. The CochleaNet contextually exploits the audio and visual cues, independent of the SNR, to estimate the spectral mask that is used to selectively suppress and enhance each time-frequency (T-F) bin. More details are comprehensively presented in \cite{m2019cochleanet}. 

 \begin{figure}[!t]
    \centering
    \includegraphics[width=1.1\linewidth]{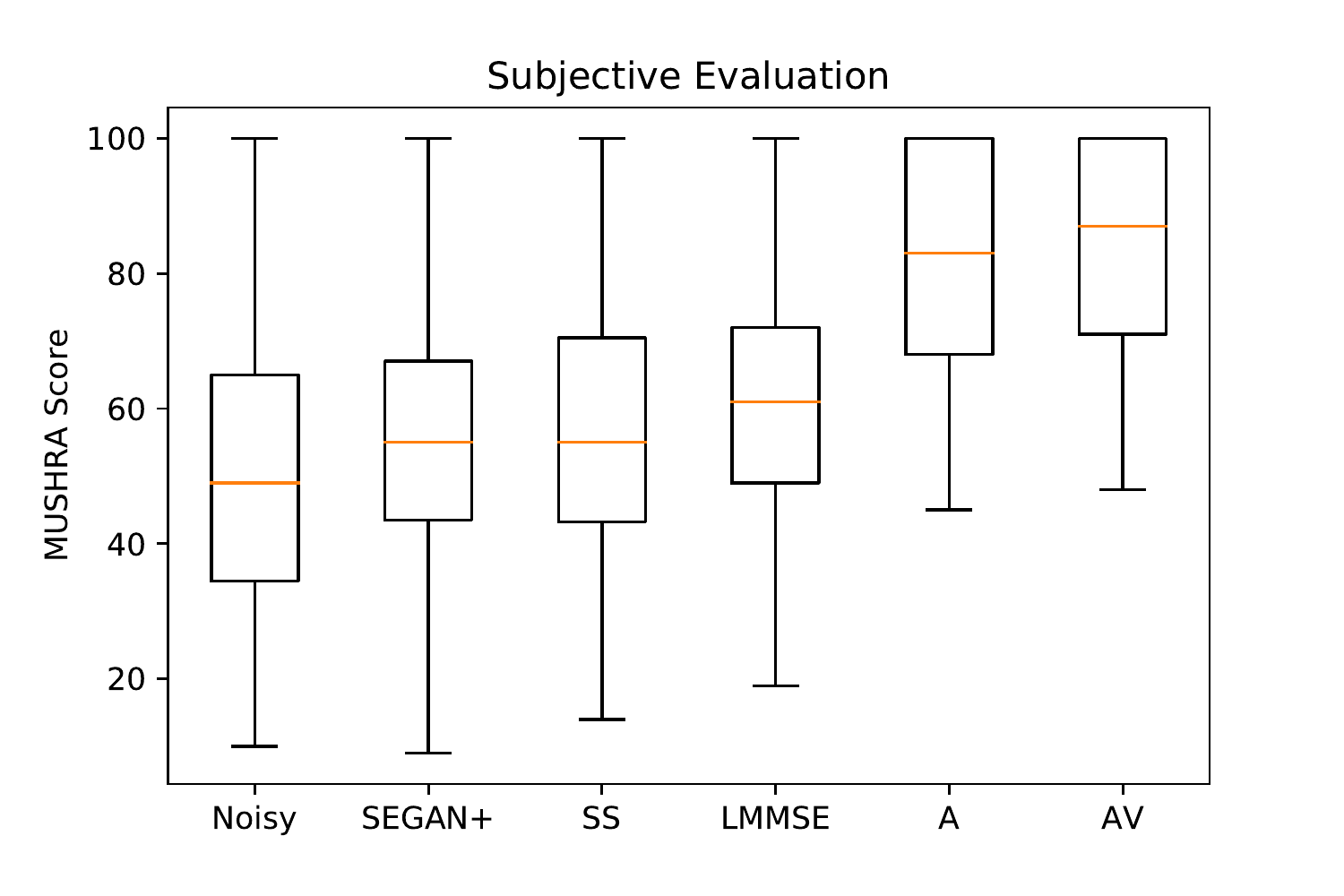}
    \caption{Result of MUSHRA listening test for ASPIRE corpus for the reconstructed speech signal using  Spectral Subtraction (SS), Linear minimum mean square error (LMMSE), SEGAN, A-only CochleaNet \cite{m2019cochleanet}, AV CochleaNet\cite{m2019cochleanet}.  The reference MUSHRA score for the unprocessed (Noisy) signal is included for relative comparison. }
    \label{fig:mushra_score}
\end{figure}

\subsection{Training}
Benchmark Grid \cite{cooke2006audio} and Chime3 \cite{barker2015third} corpora were used for the training. All 33 speakers with 1000 utterances each were considered. The sentence format is depicted in Table 1. The Grid corpus was randomly mixed with the non-stationary Chime3 noises (consisting of bus, cafeteria, street, and pedestrian noises), for SNRs ranging from -12 to 9dB, with a step size of 3 dB. It is to be noted that the trained model is SNR-independent i.e. 21000 utterances utterances at all SNRs were combined for training.  
\subsection{Testing}
MUSHRA-style \cite{recommendation20011534} listening test was used for subjective evaluation. A total of 20 native English speakers with normal-hearing participated in the listening test. The individual test consists of 20 randomly selected utterances drawn from the ASPIRE corpus. The first two screens were used to train participants, adjust volume, help participants to get familiarised with the screen and the task. Participants were asked to score the quality of each audio sample, on a scale from [0, 100], generated by different SE models for the same sentence. The range from [80, 100] is described as excellent, from [60, 80] as good, from [40, 60] as fair, from [20, 40] as poor, and from [0, 20] as bad. Noisy speech was included as a reference degraded speech. The time required to complete each screen was also recorded and used for removing any outliers. We evaluated five SE models including: SEAGN \cite{pascual2017segan}, SS \cite{boll1979spectral}, LMMSE \cite{ephraim1985speech}, A-only CochleaNet and AV CochleaNet \cite{m2019cochleanet}. Figure \ref{fig:mushra_score} shows a boxplot of listeners responses in terms of the rank order of systems for the ASPIRE corpus. Listening test results demonstrate a superior performance of AV CochleaNet, over A-only CochleaNet, SEGAN, spectral subtraction (SS), and log-minimum mean square error (LMMSE) based SE methods. Results demonstrate that the AV CochleaNet outperforms all other SE methods, generalizes well on a real-noisy corpus, and can better deal with the reverberation caused by multiple competing background sources.

\section{Conclusion}
The multimodal speech enhancement challenge is aimed at evaluating the AV speech processing systems in real-world environments. The benchmark task targets enhancement of AV binaural speech recorded in real noisy environments such as cafetaria and restaurants. The full dataset and state-of-the-art baseline have been made publicly available. A set of challenge instructions has been designed to enable meaningful comparison between benchmark approaches and maximise impact. The submitted multimodal systems and comparative results will be published at a forthcoming Interspeech Workshop. The challenge will also be extended by developing baseline tasks for Automatic Speech Recognition (ASR) communities.

\section{Acknowledgements}

This research is funded by the EPSRC project AV-COGHEAR, EP/M026981/1.



\begin{thebibliography}{10}
\providecommand{\url}[1]{#1}
\csname url@samestyle\endcsname
\providecommand{\newblock}{\relax}
\providecommand{\bibinfo}[2]{#2}
\providecommand{\BIBentrySTDinterwordspacing}{\spaceskip=0pt\relax}
\providecommand{\BIBentryALTinterwordstretchfactor}{4}
\providecommand{\BIBentryALTinterwordspacing}{\spaceskip=\fontdimen2\font plus
\BIBentryALTinterwordstretchfactor\fontdimen3\font minus
  \fontdimen4\font\relax}
\providecommand{\BIBforeignlanguage}[2]{{%
\expandafter\ifx\csname l@#1\endcsname\relax
\typeout{** WARNING: IEEEtran.bst: No hyphenation pattern has been}%
\typeout{** loaded for the language `#1'. Using the pattern for}%
\typeout{** the default language instead.}%
\else
\language=\csname l@#1\endcsname
\fi
#2}}
\providecommand{\BIBdecl}{\relax}
\BIBdecl

\bibitem{bailly2003banca}
E.~Bailly-Bailli{\'e}re, S.~Bengio, F.~Bimbot, M.~Hamouz, J.~Kittler,
  J.~Mari{\'e}thoz, J.~Matas, K.~Messer, V.~Popovici, F.~Por{\'e}e
  \emph{et~al.}, ``The {BANCA} database and evaluation protocol,'' in
  \emph{International conference on Audio-and video-based biometric person
  authentication}.\hskip 1em plus 0.5em minus 0.4em\relax Springer, 2003, pp.
  625--638.

\bibitem{lee2004avicar}
B.~Lee, M.~Hasegawa-Johnson, C.~Goudeseune, S.~Kamdar, S.~Borys, M.~Liu, and
  T.~S. Huang, ``{AVICAR}: audio-visual speech corpus in a car environment.''
  in \emph{INTERSPEECH}, 2004, pp. 2489--2492.

\bibitem{sanderson2002vidtimit}
C.~Sanderson, ``The {VidTIMIT} database,'' IDIAP, Tech. Rep., 2002.

\bibitem{cooke2006audio}
M.~Cooke, J.~Barker, S.~Cunningham, and X.~Shao, ``An audio-visual corpus for
  speech perception and automatic speech recognition,'' \emph{The Journal of
  the Acoustical Society of America}, vol. 120, no.~5, pp. 2421--2424, 2006.

\bibitem{povey2011kaldi}
D.~Povey, A.~Ghoshal, G.~Boulianne, L.~Burget, O.~Glembek, N.~Goel,
  M.~Hannemann, P.~Motlicek, Y.~Qian, P.~Schwarz \emph{et~al.}, ``The kaldi
  speech recognition toolkit,'' in \emph{IEEE 2011 workshop on automatic speech
  recognition and understanding}, no. CONF.\hskip 1em plus 0.5em minus
  0.4em\relax IEEE Signal Processing Society, 2011.

\bibitem{m2019cochleanet}
M.~Gogate, K.~Dashtipour, A.~Adeel, and A.~Hussain, ``Cochleanet: A robust
  language-independent audio-visual model for speech enhancement,''
  \emph{preprint arXiv:1909.10407}, 2019.

\bibitem{barker2015third}
J.~Barker, R.~Marxer, E.~Vincent, and S.~Watanabe, ``The third chime speech
  separation and recognition challenge: Dataset, task and baselines,'' in
  \emph{Automatic Speech Recognition and Understanding (ASRU), 2015 IEEE
  Workshop on}.\hskip 1em plus 0.5em minus 0.4em\relax IEEE, 2015, pp.
  504--511.

\bibitem{recommendation20011534}
I.~Recommendation, ``1534-1,“method for the subjective assessment of
  intermediate sound quality (mushra)”,'' \emph{International
  Telecommunications Union, Geneva, Switzerland}, 2001.

\bibitem{pascual2017segan}
S.~Pascual, A.~Bonafonte, and J.~Serra, ``Segan: Speech enhancement generative
  adversarial network,'' \emph{Interspeech}, 2017.

\bibitem{boll1979spectral}
S.~Boll, ``A spectral subtraction algorithm for suppression of acoustic noise
  in speech,'' in \emph{Acoustics, Speech, and Signal Processing, IEEE
  International Conference on ICASSP'79.}, vol.~4.\hskip 1em plus 0.5em minus
  0.4em\relax IEEE, 1979, pp. 200--203.

\bibitem{ephraim1985speech}
Y.~Ephraim and D.~Malah, ``Speech enhancement using a minimum mean-square error
  log-spectral amplitude estimator,'' \emph{IEEE Transactions on Acoustics,
  Speech, and Signal Processing}, vol.~33, no.~2, pp. 443--445, 1985.

\end{thebibliography}
\end{document}